\documentclass[12pt]{article}

\usepackage{amssymb}
\usepackage{amsmath}
\usepackage{amsfonts}

\usepackage{epsfig}

\usepackage[english]{babel}

\begin{document}

\title{\textbf{Higher Dimensions in High Energy Collisions}}
\author{\textbf{A.V. Kisselev}\\
\textit{Institute for High Energy Physics, Protvino, Russia}}

\date{}

\maketitle

\begin{abstract}
Different possibilities of a detection of signals from extra space
dimensions at high energy colliders are reviewed.
\end{abstract}


\section{Introduction}

In particle physics all the fundamental laws are formulated in
4-dimensional  Minkowski space-time. The relativistic SM fields
interact at distances $R \lesssim m_{EW}^{-1}$, where $m_{EW} \sim
10^3 \hbox{\rm \,GeV}$ is the electroweak scale. The fluctuations
of the underlying metric provide us with a gravitational dynamics.
However, the gravity becomes strong only at the Planck scale
$M_{Pl} \sim 10^{19} \hbox{\rm \,GeV}$. The first attempt to unify
gravity and particle (electromagnetic) interactions traces back to
pioneering papers by T.~Kaluza and O.~Klien~\cite{Kaluza:21}. They
have given a theoretical framework for a description of particle
interactions in a multidimentional space-time. The extra space
dimensions are strongly motivated by the string and M-theory%
~\cite{Antoniadis:90}.

The large volume of the extra dimension can explain the hierarchy
between the electroweak and Planck scale~\cite{Arkani-Hamed:98}.
The fundamental scale of the gravity may be of the order of
several TeV, if a number of additional space dimensions is large
enough. A lot of theoretical studies of an underlying theory in
the extra dimensions have been done (see, for instance,
reviews~\cite{Rubakov:01} and references therein). It is important
that physical schemes with the extra dimensions result in
distinctive phenomenological predictions which can be checked at
the LHC or at high energy linear $e^+e^-$ colliders.

In the present mini-review we consider concrete models with the
additional compactified dimensions (both with flat and warp
metric). In all of them Kaluza-Klein (KK) massive excitations of
the graviton are present which can be produced at TeV's energies.
Some of the models provide us with KK states of the SM gauge
bosons, Higgs or matter fields. The production and decays of new
scalar particles are also described. Finally, we consider a black
holes formation which can cloak hard perturbative scattering
processes behind a horizon of colliding particles.

Our main goal is a collider phenomenology. That is why we do not
consider corrections to a gravitational potential from the KK
excitations, as well as astrophysical and cosmology constraints.
Such problems as a gauge coupling unification, supersymmetry
breaking, neutrino masses and mixing, proton decay, flavor
violation \emph{etc.} are also disregarded. Because of the paper
size limitation, we present an illustrative material mainly for
the LHC, although experimental signatures of the extra dimensions
at $e^+e^-$ colliders are discussed in the text. The list of
references on the collider phenomenology within the framework of
the extra dimensions is not complete. An interested reader can
find more references in reviews~\cite{Hewett:02,Azuelos:01}.


\section{Large extra dimensions}
\label{sec:add}

\subsection{ADD model}

The large extra dimensions scenario has been postulated by
Arkani-Hamed, Dimopoulous and Dvali (ADD)~\cite{Arkani-Hamed:98}
The metric for this model looks like
\begin{equation}
ds^2 = g_{{\mu \nu}}(x) \, dx^{\mu} \, dx^{\nu} +
\eta_{_{ab}}(x,y) \, dy^a \, dy^b,
\label{add2}
\end{equation}
where $\mu,\nu = 0,1,2,3$ and $a,b=1, \ldots d$. All $d$ extra
dimensions are compactified with a characteristic size $R_c$.

There is a relation between a fundamental mass scale in $D$
dimensions, $M_D$, and 4-dimensional Planck mass, $M_{Pl}$:
\begin{equation}
M_{Pl} = V_d \, M_D^{2+d},
\label{add4}
\end{equation}
where $V_d$ is a volume of the compactified dimensions. $V_d =
(2\pi R_c)^d$ if the extra dimensions are of a toroidal form. In
order to get $M_D \sim$ TeV, the radius of the extra dimensions
should be large. The compactification scale $R_c^{-1}$ ranges from
$10^{-3}\hbox{\rm \,eV}$ to $10\hbox{\rm \,MeV}$ if $d$ runs from
2 to 6.

As we see, $R_c \gg m_{EW}^{-1}$. So, all Standard Model (SM)
gauge and matter fields are to be confined to a $3$-dimensional
brane embedded into a $(3+d)$-dimensional space (gravity alone
lives in the bulk). It means, in particular, that the
energy-momentum tensor of matter is of the form
\begin{equation}
T_{AB}(x,y) = \eta_A^{\mu}\eta_B^{\nu} T_{\mu\nu}(x)\delta(y),
\label{add6}
\end{equation}
with $A,B = 0,1, \ldots 3+d$.

In linearized quantum gravity we have
\begin{equation}
g_{_{AB}} = \eta_{_{AB}} + \frac{2}{M_D^{1+d/2}} \,
h_{_{AB}}(x,y).
\label{add8}
\end{equation}
Performing the KK mode expansion of the gravitational field
$h_{_{AB}}(x,y)$, we obtain the graviton interaction Lagrangian
\begin{equation}
\mathcal{L} = -\frac{1}{{\bar M}_{Pl}} G_{\mu \nu}^{(n)} T^{\mu
\nu},
\label{add10}
\end{equation}
where $n$ labels the KK excitation level and ${\bar M}_{Pl} =
M_{Pl}/\sqrt{8\pi}$ is a reduced Planck mass. One can conclude
from (\ref{add10}) that the coupling of both  massless and massive
graviton is universal and very small ($\sim 1/{\bar M}_{Pl}$).

The masses of the KK graviton modes are
\begin{equation}
m_n = \frac{\sqrt{n_a n^a}}{R_c} \, , \quad n_a=(n_1,n_2 \ldots
n_d).
\label{add12}
\end{equation}
So, a mass splitting is $\Delta m \sim R_c^{-1}$ and we have an
almost continuous spectrum of the gravitons.

At energies $E \gg R_c^{-1}$, the multiplicity of states which
can be produced is $N(E) \sim (ER_c)^d$. Due to a large phase
space, the cross-section of a process involving the production of
the KK gravitons with masses $m_n \leq E$ is
\begin{equation}
\sigma_{KK} \sim \frac{1}{{\bar M}_{Pl}^2} N(E) \sim
\frac{E^d}{M_D^{d+2}},
\label{add14}
\end{equation}
that is, it turns out to be measurable at future colliders.

The lifetime of the massive graviton is equal to~\cite{Han:99}
\begin{equation}
\tau_n = \frac{1}{M_{Pl}} \left( \frac{M_{Pl}}{m_n} \right)^3.
\label{add16}
\end{equation}
Thus, the KK gravitons behave like massive, almost stable
non-interacting spin-2 particles. Their collider signature is an
imbalance in missing mass of final states with a continuous mass
distribution.

\subsection{Direct production of gravitons at colliders}

\begin{figure}[ht]
\centering
\includegraphics[width=10cm,height=7cm]{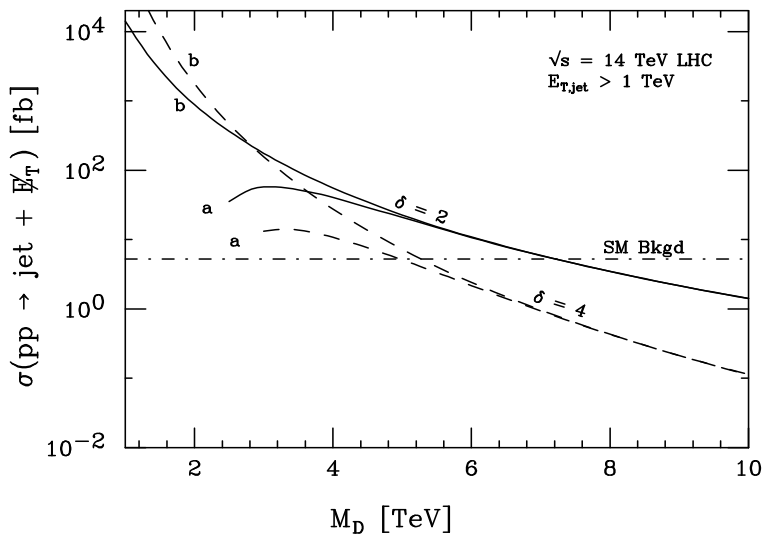}
\caption{The total ${\rm jet}+{\rm nothing}$ cross-section versus
$M_D$ at the LHC integrated for all $E_{T {\rm jet}}> 1\hbox{\rm
\,TeV}$ with the requirement that $|\eta_{\rm jet}|< 3.0$.  The SM
background is the dash-dotted line, and the signal is plotted as
solid and dashed lines for $\delta=2$ and $4$ extra dimensions.
The a (b) lines are constructed by integrating the effective
energy in the parton collision over $\hat s \leq M_D^2$ (all $\hat
s$)~\cite{Giudice:99}.} \label{fig:add}
\end{figure}

The leading experimental signal of the graviton production at the
LHC originates from the process $pp \rightarrow {\rm jet} + \not
\! \! E_T$ (the subprocess $qg \rightarrow qG^{(n)}$ gives the
largest contribution). The main background is $pp \rightarrow Z +
{\rm jet}$ ($Z \rightarrow \nu \bar \nu $). The signal and
background rates are shown in Fig. \ref{fig:add} (for the total
luminosity $\mathcal{L} = 100 \  {\rm
\,fb}^{-1}$)~\cite{Giudice:99}.

Another signal for the direct production of the massive gravitons
is the process $pp \rightarrow \gamma + \not  \! \!
E_T$~\cite{Giudice:99}. SM background comes mainly from $pp
\rightarrow Z + \gamma$ ($Z \rightarrow \nu \bar \nu $). Should
graviton be discovered in the jet channel, this process can be
used as an independent test,
 although it has the much lower rate~\cite{Giudice:99}.

The reach limit for Tevatron Run II ($E_{T {\rm jet}} > 150
\hbox{\rm \,GeV}$ and $\mathcal{L} > 300 \  {\rm \,fb}^{-1}$) was
estimated to be $M_D = 1 \hbox{\rm \,TeV}$ if $2 \leq d \leq 4$.

The graviton production can be also searched for at the future
linear colliders in the reaction $e^+e^- \rightarrow {G^{(n)}} +
\gamma /Z$, by analyzing both the total cross section and the
angular distribution of final states. SM background comes
predominantly from the process $e^+e^- \rightarrow \nu \bar \nu
\gamma$. The discovery possibility of $e^+e^-$ collider with
$\sqrt{s} = 800\hbox{\rm \,GeV}$ and $\mathcal{L} = 1000 {\rm
\,fb}^{-1}$ for different values of $d$ and beam polarizations can
be found in \cite{Giudice:99} and \cite{Hewett:02}.

Let us note that the KK excitation signature in the ADD model are
quite different from the SUSY signature. The latter is a fixed
invisible mass signal accompanied by a variety of leptons, photons
and hadron activity.

\subsection{Virtual graviton exchange}

At tree-level, the contribution of virtual massive gravitons to a
matrix element is proportional to
\begin{equation}
\mathcal{M} \sim \frac{i \pi^2}{{\bar M}_{Pl}^2} \sum_n \frac{1}{s
- m_n^2}.
\label{add18}
\end{equation}
The sum in (\ref{add18}) diverges for $d \geq 2$, the cutoff $M_H$
is to be calculated in a full theory. In the string theory it
should be related to a string scale $M_s$ and can be less than
$M_D$.

The following (rather rough) substitution is usually made for
phenomenological purposes:
\begin{equation}
\mathcal{M} \sim \frac{\lambda}{M_H^4}, \quad \lambda = \pm 1.
\label{add20}
\end{equation}

In hadron collisions the process $pp \rightarrow {G^{(n)}}
\rightarrow \gamma \gamma$ has advantages as it allows to
investigate a signal in different domains of the parton subprocess
energy. The diphoton and Drell-Yan production ($pp \rightarrow
{G^{(n)}} \rightarrow l^+ l^-$) lead to sensitivity of $M_H$ up to
$7.4\hbox{\rm \,TeV}$.

At linear colliders one of the promising reactions is $e^+e^-
\rightarrow \gamma\gamma$~\cite{Giudice:99}. The indirect effects
of the massive gravitons can be also examined in the fermion pair
production ($e^+e^- \rightarrow {G^{(n)}} \rightarrow f \bar f$).
It provides bound $M_H \lesssim 6.6\hbox{\rm \,TeV}$ (for
$\sqrt{s} = 1\hbox{\rm \,TeV}$)~\cite{Hewett:99}. The angular
distributions for the fermion pair production in $e^+e^-$
collisions (say, left-right asymmetries in $e^+e^- \rightarrow b
\bar b$) can provide a unique signature for spin-2 exchanges.
Recently a new technique was proposed which enables one to
uniquely isolate the KK gravitons from other possible new states
in the process $e^+e^- \rightarrow f \bar f$ ($f \neq
e$)~\cite{Rizzo:02}. It allows to detect the graviton exchange
contributions for mass scales up to $6 \sqrt{s}$.


\section{Warp extra dimensions}
\label{sec:rs}

\subsection{Randall-Sandrum (RS) model}

This scenario has been proposed in \cite{Randall:99}. The RS model
is a model of gravity in a 5-dimensional Anti-de Sitter space with
a single extra dimension compactified to the orbifold $S^1/Z_2$.
The metric has the form
\begin{equation}
ds^2 = e^{-2k |y|} \, \eta_{\mu \nu} \, dx^{\mu} \, dx^{\nu} +
dy^2, \label{rs2}
\end{equation}
where $y= r_c \theta $ ($0 \leqslant \theta \leqslant \pi$), $r_c$
being a "radius" of the extra dimension. The parameter $k$ defines
the scalar (negative) curvature of the space.

From the 5-dimensional action one can derive the relation
\begin{equation}
\bar M_{Pl}^2 = \frac{M_5^3}{k} \left( 1 - e^{-2kr_c\pi} \right),
\label{rs4}
\end{equation}
which means that $k \sim \bar M_5 \sim {\bar M}_{Pl}$.

There are two 3-dimensional branes in the model with equal and
opposite tension located at the point $y = \pi r_c$ (so-called the
TeV brane) and at $y = 0$ (referred to as the Plank brane). All SM
fields are constrained to the TeV brane, while gravity propagates
in the additional dimension.

Using a linear expansion of the metric
\begin{equation}
g_{\mu \nu} = e^{-2ky}\left( \eta_{\mu \nu} + \frac{2}{M_5^{3/2}}
h_{\mu \nu} \right)
\label{rs6}
\end{equation}
and a decomposition of the graviton field in KK modes, we get the
interaction of the gravitons with the SM fields
\begin{equation}
\mathcal{L} = -\frac{1}{\bar M_{Pl}} T^{\mu \nu} \, h_{\mu
\nu}^{(0)}(x) - \frac{1}{\Lambda_{\pi}} T^{\mu \nu} \, \sum_n
h_{\mu \nu}^{(n)}(x)
\label{rs8}
\end{equation}
with $\Lambda_{\pi} = {\bar M}_{Pl} e^{-kr_c \pi}$. We see that
couplings of all massive states are only suppressed by
$\Lambda_{\pi}^{-1}$, while the zero mode couples with usual
strength, $\bar M_{Pl}^{-1}$. The physical scale on the TeV brane,
$\Lambda_{\pi}$, is of the order of $1\hbox{\rm \,TeV}$ for $k r_c
\sim 12$.

The masses of the graviton KK excitations are given by
\begin{equation}
m_n = k x_n e^{-kr_c \pi},
\label{rs10}
\end{equation}
where $x_n$ are the roots of the Bessel function $J_1(x)$. It
means that the KK gravitons are not equally spaced, contrasted to
the ADD model. Thus, a basic signature of the RS model is a
resonance production.

Note that in two other RS-type models with infinitely large 5-th
dimension~\cite{Randall:99*} there is a continuum of the graviton
KK modes.

\subsection{Production of light gravitons}

For a phenomenological analysis of the RS model, the range $0.01
\lesssim k/{\bar M}_{Pl} \lesssim 0.1$ is usually used. From
theoretical considerations it follows that $\Lambda_{\pi} \lesssim
10\hbox{\rm \,TeV}$. The mass of the first excitation is expected
to be $m_1 \sim 1\hbox{\rm \,TeV}$.

The cleanest signal of the resonance production at the LHC could
be an excess in Drell-Yan process ($q\bar q,gg \rightarrow G^{(1)}
\rightarrow l^+l^-$)~\cite{Davoudiasl:00}. Notice that $gg$
initiated process now contributes to the pair production. The main
background processes are $pp \rightarrow G^{(1)} \rightarrow
Z/\gamma^* \rightarrow l^+l^-$. It was calculated that the
lightest graviton excitation would be seen if $m_1 \lesssim
2.1\hbox{\rm \,TeV}$~\cite{Allanach:00}.

The dijet production ($q\bar q,gg \rightarrow G^{(1)} \rightarrow
q\bar q,gg$) is expected to have large QCD background. Should the
first graviton resonance be discovered at the LHC, all fundamental
parameters of the model are determined through its mass $m_1$ and
total width $\Gamma_1$ by using the relations:
\begin{equation}
\Lambda_{\pi} = m_1 \frac{\bar M_{Pl}}{k x_1}, \quad \Gamma_1 =
\rho m_1 x_1^2 \left( \frac{k}{\bar M_{Pl}} \right)^2,
\label{rs12}
\end{equation}
where $\rho$ is a number of open channels for $G^{(1)}$ decay.

\begin{figure}[ht]
\centering
\includegraphics[width=7cm,height=10cm,angle=90]{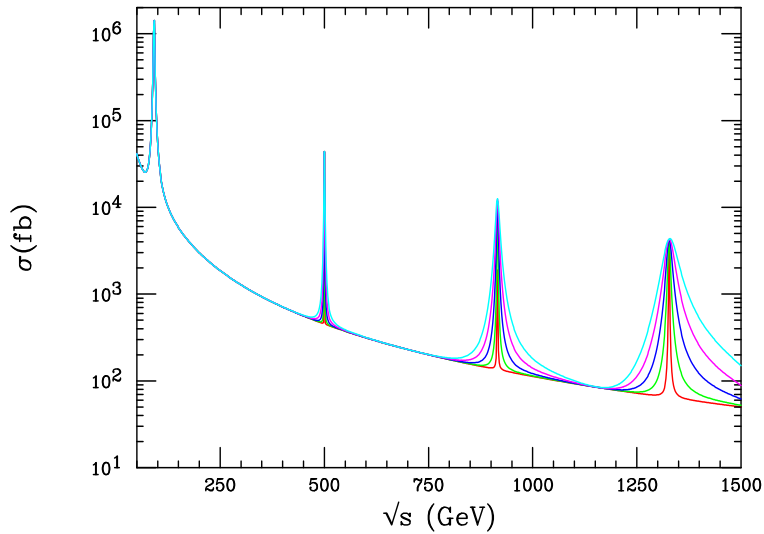}
\caption{The cross section for $e^+e^- \to \mu^+ \mu^-$ including
the exchange of a KK tower of gravitons in the Randall-Sundrum
model with $m_1=500$ GeV. The curves correspond to the range
$k/\overline M_{\rm Pl} = 0.01 - 0.05$~\cite{Hewett:02}.}
\label{fig:rs}
\end{figure}

The tower of narrow s-channel resonances can be seen in $e^+e^-
\rightarrow \mu^+ \mu^-$. The corresponding cross section as a
function of energy $\sqrt{s}$ is presented in Fig.
\ref{fig:rs}~\cite{Hewett:02}. Such resonances are easily
distinguishable from other new states ($Z'$, for instance) by
analyzing the angular distribution of the decay products.

The light graviton can be also produced in association with the
photon: $e^+e^- \rightarrow G^{(1)} + \gamma$. Contrary to large
extra dimension scheme, we expect a monoenergetic photon here.

As for virtual exchange of the massive gravitons, the
corresponding sum has no divergences. Constraints in the
$\Lambda_{\pi} - k/{\bar M}_{Pl}$ plane have been calculated in
\cite{Davoudiasl:00} from the angular distributions in
(unpolarized and polarized) $e^+e^-$ annihilation, by summing over
$e,\mu,\tau,c$ and $b$ final states.

Summary of experimental and theoretical constraints on the
parameters of RS model with the SM fields lying on the brane, can
be found in \cite{Davoudiasl:01}.


\section{\mathversion{bold} TeV$^{-1}$ Size Extra Dimensions}

In the content of the string theory it has been demonstrated that
an extra dimension may have size of the order $R_c \sim \hbox{\rm
\,TeV}^{-1}$, if the fundamental (string) scale is close to 1
$\hbox{\rm \,TeV}$~\cite{Antoniadis:90}. If the SM gauge bosons
are allowed to propagate in this compact extra dimension, their
light KK excitations have masses $m_A \sim R_c^{-1} \sim 1
\hbox{\rm \,TeV}$. As a result, the relations between electroweak
observables will be changed.

There are two kinds of effects. The first one is related to a
mixing between zero and KK modes of the $W$ and $Z$ bosons, that
results in changes of the masses of the gauge bosons and
modification of their couplings to the fermions. The second effect
is an extra contribution from a virtual exchange of the KK tower
of the gauge bosons. These effects depend on the Higgs field,
which may live in the bulk, on the walls, or can be a combination
of fields of both types~\cite{Masip:99,Rizzo:00}.

\subsection{5DSM model}

This model is an extension of the SM to the 5D flat space. The
5-th dimension $y$ is compactified on the orbifold $S^1/Z_2$ which
has two fixed points at $y = 0$ and $y = \pi R_c$. The SM gauge
fields propagate in the 5D bulk, while the chiral matter is
localized on the 4D boundaries (so-called
walls)~\cite{Pomarol:98}. The gravity can live in more extra
dimensions than the gauge fields do.

There are two Higgs doublets, $\phi_b$ and $\phi_w$, living in the
bulk and on the wall, respectively. Their vacuum expectations are
parameterized as follows
\begin{equation}
\phi_b = \upsilon \cos \beta, \qquad \phi_w = \upsilon \sin \beta,
\label{tev2}
\end{equation}
where $\upsilon$ is the standard VEV. The masses of the lightest
gauge bosons ($A = W,Z$) are~\cite{Masip:99}
\begin{equation}
m_A^{(0)2} = m_A^2 \left( 1 - 2 \sin^4 \beta \sum_{n=1}^\infty
\frac{m_A^2}{n^2 M_c^2} \right),
\label{tev4}
\end{equation}
where $M_c = R_c^{-1} \gg m_A$. The couplings of zero modes of the
$W$ and $Z$ bosons to the fermions have analogous corrections. The
KK excitations acquire the masses
\begin{equation}
M_A^{(n)2} = n^2 M_c^2 + \mbox{\rm O}(m_A^2), \quad n \geqslant 1.
\label{tev6}
\end{equation}
The coupling of the KK excitations to the SM fermions is enhanced
by a factor $\sqrt{2}$ with respect to the zero mode coupling $g$.
The more accurate calculations indicate, however, that the gauge
couplings to the matter, $g_n$, are not universal and $g_n \sim g
\exp(-A n^2/(R_cM_s)^2)$, where $M_s$ is the string  scale and $A$
is some constant~\cite{Antoniadis:94*}.

The complete analysis of the EW data gives the lower bound $M_c
\gtrsim 3.3-3.8\hbox{\rm \,TeV}$, depending on the value of $\tan
\beta$~\cite{Rizzo:00} (see also Refs.~\cite{Casalbuoni:99}).

\begin{figure}[ht]
\centering
\includegraphics[width=7cm,height=10cm,angle=90]{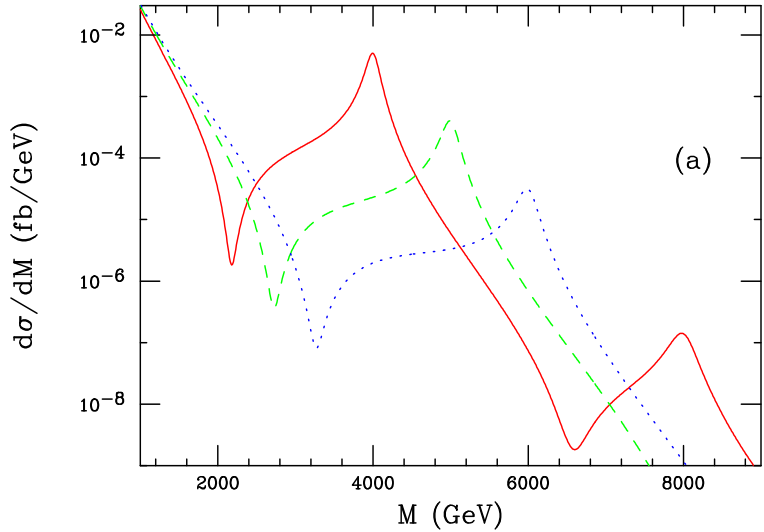}
\caption{Cross section for Drell-Yan production of the degenerate
neutral KK excitations $Z^{(n)}$ and $\gamma^{(n)}$ as a function
of the dilepton invariant mass at the LHC assuming one extra
dimension with $1/R_c$=4(5, 6) TeV corresponding to the solid
(dashed, dotted) curve. The second excitation is only shown for
the case of $1/R_c=4\hbox{\rm \,TeV}$~\cite{Rizzo:00}.}
\label{fig:tev}
\end{figure}

The states with masses not much larger than $4\hbox{\rm \,TeV}$
may be observable at the LHC. The first excitations of the gauge
bosons can be directly produced in Drell-Yan processes mediated by
the first KK modes of the gauge bosons, $pp \rightarrow
Z^{(1)}/\gamma^{(1)} \rightarrow l^+l^-$. The second level
excitations are too massive to be seen even at the LHC. The
analysis is based on a search for a narrow dilepton mass excess.
The typical cross section can be seen in Fig.~\ref{fig:tev}.

To see a direct $W^{(1)}$ production, one have to search for a
high-energy lepton plus a large missing energy. The SM background
can be significantly reduced by cuts. The reach probabilities of
the LHC are $M_c \lesssim 5.9 (6.3)\hbox{\rm \,TeV}$ for the
$Z^{(1)}/\gamma^{(1)}$ ($W^{(1)}$) channel,
respectively~\cite{Rizzo:00*} (see also estimates from
Ref.~\cite{Antoniadis:99}).

The LHC will not be able to distinguish degenerate pair of the
first excitation of $Z$ and $\gamma$ from an ordinary $Z'$ when
quarks and leptons are not at the same wall. This ambiguity can be
resolved by $e^+e^-$ collider with energy $\sqrt{s}$ much below
the masses of KK excitations. As for their values, a linear
collider with $\sqrt{s} = 1 (1.5)\hbox{\rm \,TeV}$ can probe KK
particles up to $23 (31)\hbox{\rm \,TeV}$ if integrated luminosity
$500 {\rm \,fb}^{-1}$ is obtained~\cite{Rizzo:00}. This limit is
set by the total rates and polarization asymmetries of the
processes $e^+e^- \rightarrow Z^{(n)}/\gamma^{(n)} \rightarrow f
\bar f$ for all accessible fermions.

It is interesting to note that the bulk Higgs ($\tan \beta =
\infty$) can be as large as $m_h = 500\hbox{\rm \,GeV}$ in the
5DSM, while for the wall Higgs scenario ($\tan \beta = 0$) the
value $m_h \leqslant 260\hbox{\rm \,GeV}$ is favored by the EW
data.

\subsection{RS model with bulk gauge fields}

Here we consider the case of gauge bosons propagating in the slice
of $AdS_5$ (see the previous section for details of the warp
metric). The fermions are located on the TeV brane.

In the limit $m_n \ll k$, the masses of the gauge bosons are
\begin{equation}
m_n \simeq k x_n e^{-kr_c \pi},
\label{tev8}
\end{equation}
where $x_n$ are the roots of the Bessel function
$J_0(x)$~\cite{Pomarol:00}. As was shown in~\cite{Davoudiasl:00},
the KK excitations of the gauge bosons are essentially lighter
than the graviton excitations of the same level $n$.

The ratio of the scale coupling constant of the KK mode, $g_n$ and
zero mode, $g$, is equal to
\begin{equation}
\frac{g_n}{g} = \sqrt{2\pi k r_c}, \quad n \geqslant 1.
\label{tev10}
\end{equation}
Putting $k r_c \simeq 11.27$, we get $\sqrt{2\pi k r_c} \simeq
8.4$. It means that the KK modes couple to the 3D matter fermions
about 8 times stronger than the zero mode does. Therefore, one
should expect that the tower exchange gives more significant
contribution to the EW observables that in the case with the bulk
gauge bosons living in the flat metric.

Actually, the constraint on the mass of the first excitation was
estimated to be $m_1 \gtrsim 23\hbox{\rm \,TeV}$, while
$\Lambda_{\pi} \gtrsim 100 {\rm \,TeV}$~\cite{Davoudiasl:00*}. In
Ref.~\cite{Csaki:02} lower bounds on the size of the extra
dimension, $\pi r_c$, and on the masses of the KK excitations were
calculated as functions of the Higgs mass $m_h$. In particular,
for $m_h = 115\hbox{\rm \,GeV}$ it was obtained that $(\pi
r_c)^{-1} > 11\hbox{\rm \,TeV}$, $m_1^W \gtrsim 27\hbox{\rm
\,TeV}$ and $m_1^G \gtrsim 46\hbox{\rm \,TeV}$. For the heavy
Higgs with the mass $m_h = 600\hbox{\rm \,GeV}$, the limits look
like $8.2 < (\pi r_c)^{-1} < 22\hbox{\rm \,TeV}$. Thus, one can
conclude that in the RS model with the bulk gauge fields,
resonances of both gauge bosons and gravitons lie far beyond the
reach of the the LHC.

Recently, the more complicated space-time background of the form
$AsS_5 \times \mathcal{M}^{\delta}$, where $\mathcal{M}^{\delta}$
is an orbifold with $\delta$ dimensions, was
considered~\cite{Davoudiasl:02}. For the flat orbifold $S^1/Z_2$ a
multitude of the KK graviton states appears. Moreover, couplings
of different modes are measurably non-universal, if $kR \sim 1$,
where $R$ is the radius of the $S^1$.

In another paper~\cite{Davoudiasl:02*}, the effect of boundary
kinetic terms is investigated. The substantial suppression of the
couplings of the KK gauge states was found. It means that the
lightest KK particles are of the order of a few hundred GeV and
they can be detected at the LHC contrary to the original RS model
with the bulk gauge fields. The bound on the physical mass scale
is reduced to $\Lambda_{\pi} \lesssim 10 {\rm
\,TeV}$~\cite{Davoudiasl:02*}.


\section{Universal extra dimensions (UED)}

In this section we will discuss the scheme were all the SM fields
propagate in the bulk and no walls are present. Due to the
translation invariance in the higher dimensions and conservation
of D-momentum, KK number $n$ is conserved at the tree level.
Namely, $M$ different KK modes, $n_1, n_2, \ldots n_M$ can couple
to each other if the following relation holds
\begin{equation}
|n_1 \pm n_2 \ldots \pm n_{M-1}| = n_M.
\label{ued2}
\end{equation}
The interaction of the matter fields with the gauge fields is of
the form
\begin{equation}
\mathcal{L} \sim c_{mnk} \bar f^{(m)} \gamma ^{\mu} f^{(n)}
A_{\mu}^{(k)}.
\label{ued4}
\end{equation}

Although $KK$-number is broken at one loop level, the $K$-parity,
$(-1)^n$, remains conserved~\cite{Rizzo:01}. It means that:
\begin{itemize}
    \item
    KK excitations of the SM fields must be pair-produced at
    colliders
    \item
    lowest lying KK states of light quarks and gluons are stable
\end{itemize}
As we see, the $K$-parity reminds the $R$-parity in SUSY theories.
Since the KK modes can no longer be produced as single resonances,
we have a significant reduction of the collider sensitivity to
their detection. The current experimental data put the limit
$m_{KK} \gtrsim 300-400\hbox{\rm
\,GeV}$~\cite{Apeplequist:01,Rizzo:01}.

\subsection{ACD scenario}

The starting point of this approach~\cite{Apeplequist:01} is the
minimal SM in $D = 4 + d$ flat space-time dimensions. Let us
consider the simplest case $d = 1$. In order to get chiral zero
modes of fermions, an extra dimension have to be compactified on
the orbifold $S^1/Z_2$. All other fields can be classified then as
either $Z_2$ even or odd. For instance, Higgs and gauge boson
projections on 4D space are $Z_2$ even. The orbifold only admits
the existence of the left-handed (right-handed) zero modes for the
doublet (singlet) fermion fields.

At tree level, the masses of the KK excitations comes mainly from
5D kinetic terms (with a small contribution from the Higgs):
\begin{equation}
m_n^2 = \frac{n^2}{R_c^2} + m_{SM}^2,
\label{ued6}
\end{equation}
$R_c$ being a compactification radius. Since $R_c^{-1}$ is of the
order of several hundred $\hbox{\rm \,GeV}$, the KK states of the
same level are almost degenerate in their masses. The real
signature of the UED is a production of large number of heavy
stable particles.

\begin{figure}[ht]
\centering
\includegraphics[width=7cm,height=10cm,angle=90]{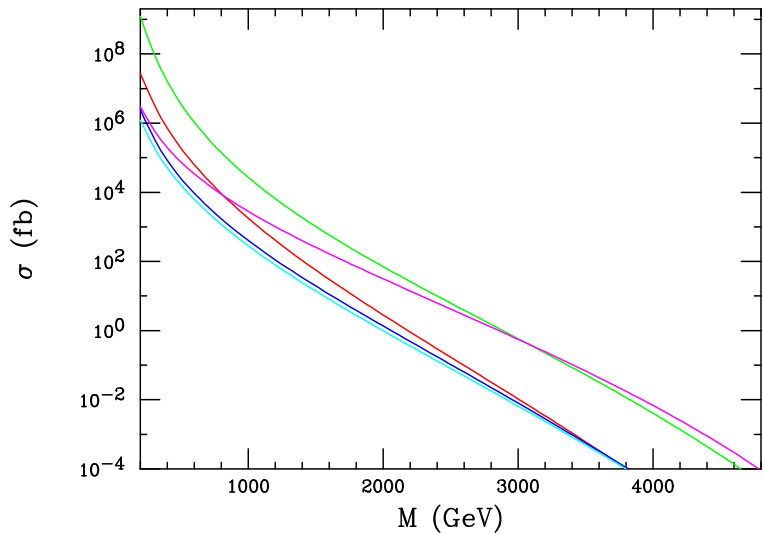}
\caption{Cross section for the pair production of the lightest
colored KK states at the LHC~\cite{Rizzo:01}. From top to bottom
on the left-hand side, the curves correspond to the processes
given in (\ref{ued8}).}
\label{fig:ued}
\end{figure}

There are three classes of processes to be investigated at hadron
colliders. The processes with color final states have the largest
cross sections. they are mediated by the subprocess:
\begin{eqnarray}
qq' &\rightarrow& q^{(1)} q^{'(1)}, \nonumber\\
q \bar q &\rightarrow& q^{(1)} \bar q^{(1)},  \nonumber\\
gg + q \bar q &\rightarrow& q^{'(1)} \bar q^{'(1)}, \nonumber\\
gg &\rightarrow& g^{(1)} g^{(1)}, \nonumber \\
qq &\rightarrow& q^{(1)} q^{(1)}.
\label{ued8}
\end{eqnarray}
The most significant subprocess is $qq' \rightarrow q^{(1)}
q^{'(1)}$. The LHC will be able to probe the masses $m_{KK}
\lesssim 3\hbox{\rm \,TeV}$~\cite{Rizzo:01}. The cross sections of
the pair production of the lightest colored states are presented
in Fig.~\ref{fig:ued}.

For the associated production of the lightest color singlet KK
states ($g^{(1)}W^{(1)}$, $g^{(1)} Z^{(1)}$, $g^{(1)}
\gamma^{(1)}$, $q^{(1)}W^{(1)}$, $q^{(1)} Z^{(1)}$, $q^{(1)}
\gamma^{(1)}$), masses up to $1.5\hbox{\rm \,TeV}$ can be
detected. Finally, the production of color singlet states
($2Z^{(1)}Z^{(1)}$, $\gamma^{(1)} Z^{(1)}$, $2\gamma^{(1)}$,
$W^{(1)+}W^{(1)-}$, $W^{(1)\pm}Z^{(1)}$, $W^{(1)\pm}\gamma^{(1)}$)
will be probed at the LHC if $m_{KK} \lesssim 1.8\hbox{\rm
\,TeV}$~\cite{Rizzo:01}. The limit on $R_c^{-1}$ accessible on the
LHC was estimated to be $13.5-15.5\hbox{\rm
\,TeV}$~\cite{Azuelos:01}.

At future $e^+e^-$ colliders the KK excitation of the SM fields
can be discovered in the processes $e^+e^- \rightarrow
W^{(1)}W^{(1)}, \, 2\gamma^{(1)},\, Z^{(1)}\gamma^{(1)},\,
2Z^{(1)}$.

If the $K$-parity is conserved, the production of stable heavy KK
particles will be missed at colliders. However, it can appear that
a new physics makes them unstable. For instance, if the UED is
embedded in a higher $(4 + d')$-space with $d' > d$ and
compactification radius $R_c' > R_c$, there are transitions of the
form $KK(n=1) \rightarrow KK(n=0) + G$. Another possible way of
violating the $K$-parity is to introduce an additional brane. The
mixing between all KK excitations and zero mode results in then
the transition $KK(n=1) \rightarrow 2 \, KK(n=0)$, and the KK
states can decay inside a detector (see \cite{Apeplequist:01} and
\cite{Rizzo:01} for details).

\subsection{Bulk fermions in RS background}

As we already know, the KK excitations of the gauge fields are
significantly lighter than the corresponding KK excitations of the
graviton. In Ref.~\cite{Chang:00} the masses of the bulk fermions
were found to be $m_n^f = n \pi k /[\exp(\pi k r_c) - 1]$.

In a consequent paper~\cite{Davoudiasl:01}, the Dirac 5D mass term
required by the $Z_2$ orbifold symmetry was introduced in the
action. For simplicity all fermions are taken to have the same
mass $m = \nu k$, where $\nu$ is an arbitrary parameter. In such a
scheme the masses of the KK excitations have an approximately
linear dependence on $\nu$:
\begin{equation}
m_n^f = a_n \left|\nu + 1/2 \right| + b_n,
\label{ued14}
\end{equation}
$a_n,\,b_n$ being some constants. The KK fermion states are
expected to be more massive than the corresponding KK states of
the gauge bosons~\cite{Davoudiasl:01}. In the rest of this section
we consider this model.

The couplings of the resulting KK states strongly depend on the
value of $\nu$. For instance, the ratio $g_n/g$ is quite small for
$\nu \lesssim -0.5$, while for $\nu \gg 1$ the result for the RS
approach with the wall fermions (\ref{tev10}) is reproduced.

The phenomenology of the model is determined by the parameters
$k,\,\Lambda_{\pi}$ and $\nu$. The primary discovery mode for the
gauge bosons at hadron colliders could be Drell-Yan process ($p
\bar p \rightarrow \gamma^{(1)}/Z^{(1)} \rightarrow l^+l^-$), but
it gives no new signature. Unfortunately, both the
$W^+W^-$--fusion and the top production are also poor places to
search for KK events~\cite{Hewett:02*}. Thus, it seems unlikely
that the KK excitations will be produced directly at the
LHC~\cite{Hewett:02*, Huber:02}.

As for future linear colliders, the range $m_1 \lesssim
15\hbox{\rm \,TeV}$ can be probed for $\nu \lesssim -0.3$ in
$e^+e^- \rightarrow \gamma^{(1)}/Z^{(1)} \rightarrow f \bar f$ ($f
= e,\,\mu, \ldots t$), if center-of-mass energy $\sqrt{s} =
500-1000\hbox{\rm \,GeV}$ and integrated luminosity $\mathcal{L} =
500-1000 {\rm \,fb}^{-1}$ are obtained~\cite{Hewett:02*}.


\section{Minimal universal extra dimension (MUED)}

This model proposed in Refs.~\cite{Cheng:02} is defined in five
dimensions with an additional dimension compactified on the
$S^1/Z_2$ as in the UED. The full Lagrangian of the MUED contains
both bulk and boundary interactions. The latter are localized at
the fixed points of the orbifold, and, thus, do not respect 5D
Lorenz invariance. The boundary terms are chosen to be symmetric
under the exchange of the orbifold fixed points, and the
$K$-parity remains an exact symmetry.

These new interactions:
\begin{itemize}
    \item
    violate the $KK$ number by even units
    \item
    lead to a mass splitting between the KK modes
    \item
    affect their decays
\end{itemize}
The boundary term are assumed to be small at some scale $\Lambda
> R_c^{-1}$.

There are only three free parameters in the MUED: $R_c,\,\Lambda$
and $m_h$, where $m_h$ is the mass of the SM Higgs. The dependence
of the splitting between KK excitation with $n = 1$ on the scale
$\Lambda$ (at fixed $R_c^{-1} = 500\hbox{\rm \,GeV}$ and $m_h =
120\hbox{\rm \,GeV}$) has been calculated in~\cite{Cheng:02}. The
corrections to the masses are such that the heaviest (lightest)
particle is the first KK states of the gluon (photon). The $SU(2)$
doublet quarks, $Q^{(1)}$, are heavier that singlet quarks,
$q^{(1)}$.

\begin{figure}[ht]
\centering
\includegraphics[width=10cm,height=7cm]{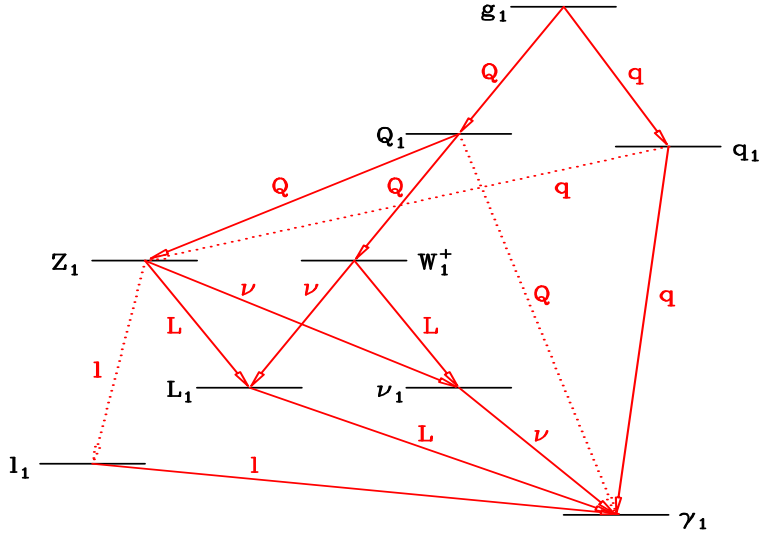}
\caption{Qualitative sketch of the level 1 KK spectroscopy
depicting the dominant (solid) and rare (dotted) transitions and
the resulting decay product~\cite{Cheng:02}.}
\label{fig:mued}
\end{figure}

The collider phenomenology of the first KK level is very similar
to the SUSY scheme in which the superpartners are close in their
masses (with $\Delta m = 100-200\hbox{\rm \,GeV}$). Like the LSP
is stable in $R$-parity conserving SUSY, the lightest KK particle
(LKP), $\gamma^{(1)}$, is stable due to the $K$-parity
conservation. The KK spectroscopy and possible transitions in the
MUED is shown in Fig.~\ref{fig:mued}.

At hadron coliders the KK states are pair-produced. The process
with the largest rate is $pp \rightarrow N {\rm jet} + \not  \! \!
E$ ($N \geqslant 2$). It is mediated by an inclusive production of
$q^{(1)} q^{(1)}$ and it is similar to the usual squark (sgluino)
search. The measured missing energy is rather small. The LHC is
expected to reach only mass scales up to $R_c^{-1} \lesssim
1.2\hbox{\rm \,TeV}$~\cite{Cheng:02}. The more promising process
is an inclusive production of $Q^{(1)}Q^{(1)}$-pair with the
subsequent decays of the quark KK excitations into leptonic final
states. As a result, we have a signature $4l + \not  \! \! E$. The
main background comes from the subprocess $Z(\gamma)Z(\gamma)
\rightarrow l^{\pm} l^{\mp} \tau^+ \tau^- \rightarrow 4l + \not \!
\! E$ ($Z$ and $\gamma$ may be real or virtual). The scale
$R_c^{-1} \sim 1.5\hbox{\rm \,TeV}$ can be reached. Another
channels with two or three leptons in the final state have larger
backgrounds.

We have already seen that MUED signals may be easily confused with
the SUSY signals. It is a linear collider that can help us to
distinguish the MUED from SUSY. One such a possibility is the
production of KK electrons in $e^-e^-$ collisions~\cite{Cheng:01}.
As for the LHC, one has to look for the existence of the KK modes
with $n \geqslant 2$ which should be the smoking gun signature for
the MUED.


\section{Radion}

\subsection{Graviscalar-Higgs mixing}

Let us analyze scalar content of the ADD scheme with $d$
additional compact flat dimensions which was considered in
Section~\ref{sec:add}. The KK tower of a 5D gravitational field
$h_{AB}$ ($A,B = 0,1, \ldots 4+d$) contains a set of
gauge-invariant fields of different spin: $G_{\mu\nu}^{(n)}$,
$V_{\mu i}^{(n)}$, $S_{ij}^{(n)}$ and $H^{(n)}$, where indexes
$i,j$ run over extra dimensions. From the point of view of four
dimensions, $G_{\mu\nu}^{(n)}$ is the massive spin-two particle
($n$-th KK excitation of the graviton). Both vector fields,
$V_{\mu i}^{(n)}$, and massive real scalars, $S_{ij}^{(n)}$, are
not coupled to the matter and play no roles in a phenomenology.
The scalars $H^{(n)}$ are coupled only to the trace of the
energy-momentum tensor~\cite{Giudice:99}:
\begin{equation}
\mathcal{L}_{int} = - \frac{\kappa}{3 \bar M_{Pl}} \sum_n H^{(n)}
T_{\mu}^{\mu},
\label{rad2}
\end{equation}
where $\kappa = \sqrt{3(d - 1)/(d + 2)}$.

For $d = 1$, only zero massless mode of $H^{(n)}$, called radion,
is present, while KK excitations of the radion are eaten by the
massive gravitons. The radion corresponds to fluctuations of the
volume of the extra dimension. There exist several mechanisms
which stabilize the radius of the extra dimension and give a mass
to the radion (see, for instance, \cite{Arkani-Hamed:01}). At $d
> 1$, there is a tower of massive graviscalars.

At tree level the coupling of the scalars to the massless fermions
and gauge bosons vanishes, as one can see from (\ref{rad2}).
However, the graviscalars may play an important role if there is a
scalar fields, $\varphi$, living on the brane. The point is that
one can add the following 4D term, involving the Ricci scalar $R$,
to 4D effective Lagrangian
\begin{equation}
\Delta \mathcal{L} = - \xi R \varphi^ + \varphi,
\label{rad4}
\end{equation}
with a free parameter $\xi$. In the unitary gauge, $\varphi$ is
reduced to the physical Higgs via the relation $\varphi = (v + h,
\, 0)/\sqrt{2}$, $\upsilon$ being the VEV of the Higgs field.

Taking into account (\ref{rad2}) and (\ref{rad4}), we get
graviscalar-Higgs mixing term of the form~\cite{Giudice:01}
\begin{equation}
\mathcal{L}_{mix} = \frac{2\kappa \xi \upsilon m_h^2}{{\bar
M}_{Pl}} \, h \sum_n H^{(n)}.
\label{rad6}
\end{equation}

As it has been shown in~\cite{Giudice:01}, an oscillation of the
Higgs field into a huge number ($\sim {\bar M}_{Pl}^2$) of closely
spaced scalars is equivalent to a decay of the Higgs on collider
time scale with partial width
\begin{equation}
\Gamma \sim \pi \kappa \, \xi^2 \upsilon^2
\frac{m_h^d}{M_D^{2+d}}.
\label{rad8}
\end{equation}
This width corresponds to the invisible decay of the Higgs, since
the graviscalars will escape a detector.

The search for the invisible decay at the LHC could help us to
discover the Higgs provided its mass is below $250
\hbox{\rm\,GeV}$. At $e^+e^-$ collider with the energy $\sqrt{s} =
500 \hbox{\rm\,GeV}$ and integrated luminosity $\mathcal{L} = 200
{\rm \,fb}^{-1}$, the range $80 < m_h < 170 \hbox{\rm\,GeV}$ can
be probed~\cite{Giudice:01}.

\subsection{Radion in RS model}

In the RS model (see Section \ref{sec:rs}) the radion field arises
as an excitation of the volume of the slice of the $AdS_5$ space,
$r_c \rightarrow T(x)$. In terms of the dynamical field $T(x)$,
the metric of the model (\ref{rs2}) may be presented in the form
\begin{equation}
ds^2 = e^{-2 k |\theta| T(x)} \eta_{\mu \nu} \, dx^{\mu} \,
dx^{\nu} + T^2(x) d\theta^2. \label{rad10}
\end{equation}
Let us introduce more conventional definition $\phi =
\Lambda_{\phi} e^{-k \pi (T - r_c)}$. Here $\Lambda_{\phi}
 = \sqrt{24M_5^3/k}\,  \exp(-k \pi r_c)$ and
$M_5$ is the 5D Planck scale. Due to a mechanism stabilizing the
size of the extra dimension (see, for instance,
\cite{Goldberger:99,Csaki:00}), the radion $\phi$ acquires a
dynamical mass of the order of $1 \hbox{\rm \,TeV}$ or less. The
common expectation is that the radion may be the lightest new
particle in the RS model.

The interaction of the radion $\phi$ with the SM fields located on
the brane is given by
\begin{equation}
\mathcal{L}_{int} = - \frac{\phi}{\Lambda_{\phi}} T_{\mu}^{\mu}.
\label{rad12}
\end{equation}
The radion phenomenology is very similar to the SM Higgs up to the
rescaling factor $v/\lambda_{\phi}$, except that its couplings to
two photons and gluons are enhanced by the trace anomaly. The
$\phi \phi h$ coupling can be substantially larger than the $h h
h$ coupling~\cite{Bae:00}.

Relatively light radion decays dominantly into two gluons (the
decay is enhanced by the gluon anomaly presented in
$T_{\mu}^{\mu}$) contrary to the SM Higgs which decays into $b
\bar b$ pair. If the radion mass lies within the limits $150
\lesssim m_{\phi} \lesssim 2m_W$, the main mode will be $\phi
\rightarrow hh$ (if kinematically allowed). The heavier radion
decays into $W^+W^-$ and $ZZ$.

\begin{figure}[ht]
\centering
\includegraphics[width=10cm,height=7cm]{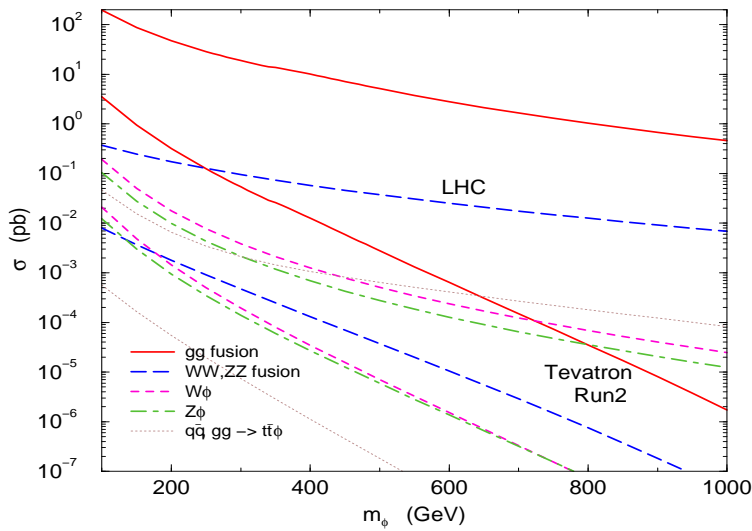}
\caption{Production cross sections versus $m_\phi$ for $p\bar p
\to \phi$ ($gg$ fusion), $p\bar p \to q q' \phi$ ($WW,ZZ$ fusion),
$p\bar p \to W\phi$, $p\bar p \to Z\phi$, and $p\bar p \to t \bar
t \phi$~\cite{Cheung:01}.}
\label{fig:rad}
\end{figure}

Let us consider first the detection of the radion when it does not
mix with the Higgs. At hadron colliders the production
subprocesses are $gg \rightarrow \phi$ (the most important
channel), $q \bar q' \rightarrow W \phi$, $q \bar q \rightarrow Z
\phi$, $q q' \rightarrow q q' \phi$ and $q \bar q, \, gg
\rightarrow t \bar t \phi$~\cite{Cheung:01}. The cross section can
be seen in Fig.~\ref{fig:rad}.

The dijet production ($gg \rightarrow \phi \rightarrow gg$) has
too large background and a signal is about $1\%$ at the LHC. The
$b \bar b$ mode is not also a good place for a radion detection,
as the corresponding branching is ten times smaller than that of
the SM Higgs. For rather heavy radion ($m_{\phi} > 180 {\rm
\,GeV}$) the cleanest signature is~\cite{Giudice:01,Cheung:01}
\begin{equation}
gg \rightarrow \phi \rightarrow ZZ \rightarrow 4l.
\label{rad14}
\end{equation}
The radion has a smaller width than the Higgs. Thus, one has to
search for a small invariant mass of four leptons in
(\ref{rad14}). The LHC can reach $m_{\phi} \lesssim 1 {\rm
\,GeV}$. In the $WW$ mode ($gg \rightarrow \phi \rightarrow WW$
$\rightarrow l\bar l \nu \bar \nu$) the radion can be defined in
the mass range $m_{\phi} = 140-190 {\rm \,GeV}$ if $\Lambda_{\phi}
\simeq \upsilon$~\cite{Cheung:01}.

At linear colliders, the radion can be produced in $e^+e^-
\rightarrow Z \phi$, $\nu \bar \nu \phi$, $e^+e^- \phi$. The
search strategy should be similar to that of the Higgs, but with
the detection of two jets instead of $b \bar b$-pair in final
states.

The radion-Higgs mixing induced by the term (\ref{rad4}) shifts
the properties of both fields~\cite{Giudice:01,Rizzo:02*} which
become dependent on a mixing parameter $\xi$. For instance, the
branching fraction $h \rightarrow gg$ can be strongly modified
with respect to its SM expectation~\cite{Rizzo:02**}. The case
$\xi = 1/6$ ("half Higgs, half radion") is the special one. The
coupling of $\phi$ to the fermions, $W$ and $Z$ are suppressed for
$\xi = 1/6$ and the $gg$-branching of the radion dominates for all
values of $m_{\phi}$~\cite{Giudice:01}. In Ref.~\cite{Giudice:01}
the search capability of the LHC has been estimated:
\begin{itemize}
    \item
    $110 < m_{\phi} \leq 150 \mbox{\rm \,GeV}$
    ($2 \lesssim \Lambda_{\phi} \lesssim 3 \mbox{\rm \,TeV}$)
    \item
    $150 < m_{\phi} \leq 550 \mbox{\rm \,GeV}$
    ($3 \lesssim \Lambda_{\phi} \lesssim 7 \mbox{\rm \,TeV}$)
    \item
    $550 < m_{\phi} \leq 950 \mbox{\rm GeV}$
    ($4 \lesssim \Lambda_{\phi} \lesssim 7 \mbox{\rm \,TeV}$)
\end{itemize}

The radion properties in the RS scenario where the SM field
propagate in the bulk, were considered in~\cite{Rizzo:02*}. It was
shown that there is no differences in $f \bar f$, $hh$ modes, but
massless modes are seriously modified. Namely, the width
$\Gamma(\phi\rightarrow gg)$ receives $40-50\%$ increase and the
width $\Gamma(\phi\rightarrow gg)$ changes even more drastically.


\section{Black holes at colliders}

\subsection{Black hole production}

As we know from Section \ref{sec:add}, the 5D Planck scale, at
which the gravity becomes strong, is $M_D \sim 1 \hbox{\rm\,TeV}$
for $D = 10$. If this scenario is realized in nature, a production
of black holes should be possible at super-Planckian energies
($\sqrt{s} \gg 1 \hbox{\rm\,TeV}$). Black hole intermediate states
are expected to dominate $s$-channel scattering. Indeed, in the
string theory the number of such states grows with black hole
mass, $M_{BH}$, faster than the number of perturbative
states~\cite{Giddings:02}. Moreover, as we will see below, the
cross section of the black hole production rises in $\sqrt{s}$
more rapidly than in processes associated with the perturbative
physics.

The Schwartzchild radius of a $(4 + d)$ dimensional black hole
with the mass $M_{BH}$ depends on its spin $J$. For $J = 0$ it is
given by~\cite{Myers:86}
\begin{equation}
R_S(M_{BH}) = \frac{1}{M_D} \left( \frac{M_{BH}}{M_D}
\right)^{1/(1+d)}. \label{bh2}
\end{equation}
In what follows, it is assumed that
\begin{equation}
R_S \ll R_c \label{bh4},
\end{equation}
where $R_c$ is a characteristic geometrical scale of the model. In
addition, the black hole masses larger than the tension are only
considered. These two assumptions mean that we can use flat space
formulae.

Then the black hole can be considered as a neutral spinning
solution of the $D$ dimensional Einstein action with the radius
$R_S$ (\ref{bh2}), the Hawking temperature
\begin{equation}
T_H \xrightarrow{J \rightarrow 0} \frac{d+1}{4\pi R_S} \label{bh6}
\end{equation}
and the entropy
\begin{equation}
S_{BH}  \xrightarrow{J \rightarrow 0} {\rm const} \left( R_S M_D
\right)^{d + 2}, \label{bh8}
\end{equation}
where the constant in the RHS of Eq.~(\ref{bh8}) depends on $d$.

It is necessary to define the applicability of the description of
the black hole as a massive semi-classical state. The main quantum
corrections are due to the change in the Hawking
temperature~(\ref{bh6}) per particle emission. The condition that
back-reaction of this emission on the black hole to be small is
equivalent to $S_{BH} \gg 1$. Another criteria is the validity of
the statistical description for the black hole, that results in
the inequality $\sqrt{S_{BH}} \gg 1$~\cite{Giddings:02}. The value
of $S_{BH} \gtrsim 25$ is usually taken, which means $M_{BH}
\gtrsim 5 M_D$. As for classical modifications of gravity (string
corrections), the corresponding effects can be neglected if $R_S >
M_s$, where $M_s$ is the string scale.

Let us consider two colliding objects with the energy $\sqrt{s}$.
If an impact parameter $b$ becomes smaller than some critical
value (which is of the order of the Schwartzchild radius
$R_S(s)$~(\ref{bh2})), the black hole is formed. As $R_S(s)$ is
large and grows with energy, the production of the black hole can
be described by classical General Relativity~\cite{Bank:99}.

So, at $b \leqslant R_S(s)$, the scattering is dominated by
``resonant'' production of a single black hole with $M_{BH} =
\sqrt{s}$. It is important to note that the event horizon of
colliding particles forms long before their collision takes place.
In hadron-hadron scattering the cross section of the black hole
production is of the form
\begin{equation}
\sigma_{pp \rightarrow BH}(s) = \sum_{a,b} \int_{\tau_{min}}^1 \!
d \tau \int_{\tau}^1 \frac{dx_a}{x_a} f_a(x_a)
f_b(\frac{\tau}{x_a}) \sigma_{ab \rightarrow BH}(\tau s).
\label{bh10}
\end{equation}
Here $\tau_{min} = M_{BH}^{min}/s$ and the black hole mass is
assumed to be $M_{BH} \simeq \sqrt{\tau s}$. Following Thorne's
hoop conjecture, the cross section of two partons, $a$ and $b$,
are taken in a simple geometric form~\cite{Giddings:02}:
\begin{equation}
\sigma_{ab \rightarrow BH}(s) \approx \pi R_S^2(s).
\label{bh12}
\end{equation}
Recent calculations (see, for instance, \cite{Eardley:02})
indicate that quantum corrections to the semi-classical
expression~(\ref{bh12}) are small.

As we can see, the cross section has no small coupling constants
and it rises rapidly with the energy, while hard perturbative
processes are highly suppressed above the scale $M_D$. The
summation over all types of initial partons in (\ref{bh10})
results in an additional enhancement of the black hole production.
Once the black hole is formed, colliding particles never get close
enough to perform a hard scattering. With TeV scale gravity, the
production of the black holes should be a dominant process at the
LHC. The total cross section can be as large as $0.5 \,(120) {\rm
\,fb}$ for $M_{BH} = 2 \,(6) \hbox{\rm\,TeV}$ and $d = 7
\,(3)$~\cite{Dimopoulos:01}.

\subsection{Black hole decays}

The black hole decay occurs in several stages. The fist stage is
so-called balding phase, where highly asymmetrical black hole
performs classical gauge and gravitational radiation. As a result,
the black hole with no hair is produced. One expects that the
black hole emits $16\%$ of its mass.

The black hole is formed with rather large  spin $J \sim R_S
M_{BH}$. In spin-down phase it is shed in quanta with angular
momentum $L \sim 1$ and energy $E \sim R_S^{-1}$. About $25\%$ of
the black hole's energy is radiated during this phase.

The remaining energy of the black hole is mainly lost in a
Schwartzchild phase. At this stage, the black hole's evaporation
is a thermal (grey body) emission at Hawking temperature
(\ref{bh6}). It is important that the black hole decays visibly to
the SM particles living on the brane~\cite{Emparan:00}. The total
number of emitted particles is equal to $\langle N \rangle \simeq
M_{BH}/T_H$. The typical black hole's lifetime is $\tau_{BH} \sim
10^{-26} {\rm \,sec}$, which corresponds to the total width
$\Gamma_{BH} \sim 10 {\rm \,GeV}$~\cite{Dimopoulos:01}.

Finally, the black hole with a mass $M_{BH} \sim M_D$ decays into
several quanta with energies ${\rm O}(M_D)$. This decay is called
Planck phase.

The experimental signatures of the black hole production are very
distinctive:
\begin{itemize}
    \item
    flavor-blind (thermal) decays
    \item
    hard prompt charged leptons and photons (with energy $E
    \gtrsim 100 {\rm \,GeV}$)
    \item
    the ratio of hadronic to leptonic activity is closed to
    $5:1$~\cite{Giddings:02}
    \item
    complete cut-off of hadronic jets with $p_{\perp} > R_S^{-1}$%
    ~\cite{Bank:99}
    \item
    the small missing energy
\end{itemize}

\begin{figure}[ht]
\centering
\includegraphics[width=10cm,height=7cm]{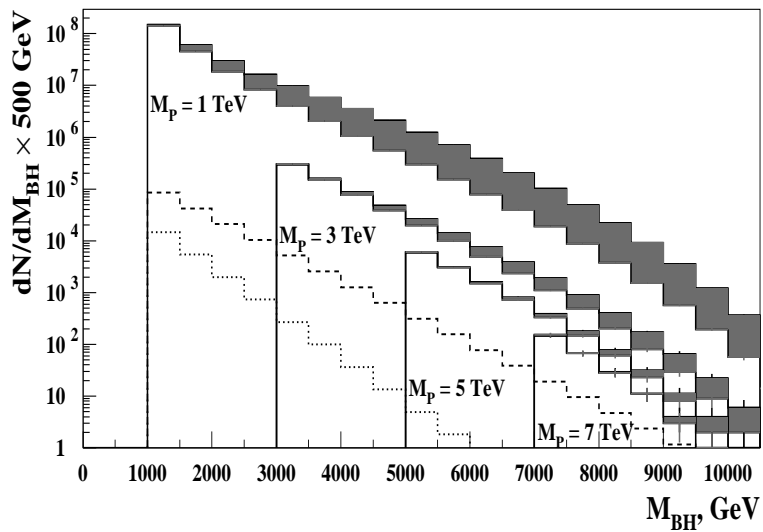}
\caption{Number of black holes produced at the LHC in the electron
or photon decay channels, with $\mathcal{L} = 100 \,{\rm
\,fb}^{-1}$, as a function of $M_{BH}$. The shaded regions
correspond to the variation in the number of events for $d$
between 2 and 7. The dashed line shows total SM background (from
inclusive $Z(ee)$ and direct photon production). The dotted line
corresponds to the $Z(ee) + X$ background
alone~\cite{Dimopoulos:01}.}
\label{fig:bh}
\end{figure}

These signatures (considered together) have almost vanishing SM
background. The LHC discovery potential is maximized in $e/\mu +
X$ channel. The multiplicity distribution of the black holes
produced is presented in Fig.~\ref{fig:bh}. The SM backgrounds
from $Z(e^+e^-) + {\rm jet}$ or $\gamma + {\rm jet}$ final states
are small and scales up to $M_D \lesssim 9 \hbox{\rm\,TeV}$ can be
reached~\cite{Dimopoulos:01}. The production cross sections of
black holes were also calculated in \cite{Cheung:02}.

If the dependence of the Hawking temperature $T_H$ vs. $M_{BH}$
will be measured, one can determine the number of the extra
dimensions by taking the logarithm of both sides of
Eq.~(\ref{bh6}):
\begin{equation}
\log T_H  = - \frac{1}{d+1} \log M_{BH} + {\rm const},
\label{bh14}
\end{equation}
where the constant depends only on $M_D$, but not on $d$.

However, the evaporation of the black hole may significantly
change in models with separated fermions, once $R_S$ is smaller
than the separation distance~\cite{Han:02}. Then the ratio of jets
to charged leptons and photons becomes $113:8:1$. The spin of the
black hole also affects energy and angular spectra of the Hawking
radiation~\cite{Kotwal:02}. Moreover, some authors argue that
$D$-dimensional black hole must radiate mainly into KK
modes~\cite{Argyres:98,Bank:99}. Note that the key geometric
formula (\ref{bh12}) needs more justification as well.


\section{Conclusions}

The LHC will be able to detect signals from higher-space
dimensions and to reach the effective Planck scale up to $5-10
{\rm \,TeV}$, depending on the model and the experimental
signature. The $e^+e^-$ collisions should be effective in the
indirect search for the KK excitations. In the scheme with
non-factorizable metric, the masses of KK states as large as
several ten TeV can be probed at the linear collider with energy
$\sqrt{s} = 1{\rm \,TeV}$ and integrated luminosity $\mathcal{L} =
500 {{\rm \,fb}^{-1}}$.


\section*{Acknowledgments}

I am grateful to the Organizing Committee of the Workshop for
suggesting me a subject for my review talk and to V.A. Petrov for
stimulating discussions.


\end{document}